\documentstyle[prl,aps,epsfig]{revtex}
\begin{document} 
\draft
\twocolumn[\hsize\textwidth\columnwidth\hsize\csname@twocolumnfalse%
\endcsname
\preprint{\parbox[t]{45mm}{\small ANL-PHY-8683-TH-97 \\ UNITU-THEP-7/1997\\  
TU-GK-97-002\\ hep-ph/9705242 \\}}
 
\title{The Infrared Behavior of Gluon and Ghost Propagators in 
Landau Gauge QCD}
\author{Lorenz von Smekal}
\address{Physics Division, Argonne National Laboratory,
         Argonne, Illinois 60439}
\author{Andreas Hauck and Reinhard Alkofer}
\address{Institut f\"{u}r Theoretische Physik,
         Universit\"{a}t T\"{u}bingen,
         Auf der Morgenstelle 14, 72076 T\"{u}bingen, Germany}
\date{\today}
\maketitle
\begin{abstract}
A solvable systematic truncation scheme for the Dyson--Schwinger equations
of Euclidean QCD in Landau gauge is presented. It implements the
Slavnov--Taylor identities for the three--gluon and ghost--gluon vertices,
whereas irreducible four--gluon couplings as well as the gluon--ghost and
ghost--ghost scattering kernels are neglected. The infrared behavior of gluon
and ghost propagators is obtained analytically: The gluon propagator vanishes
for small spacelike momenta whereas the ghost propagator diverges more
strongly than a massless particle pole. The numerical solutions are compared
with recent lattice data for these propagators. The running coupling of the
renormalization scheme approaches a fixed point, $\alpha_c \simeq 9.5$, in
the infrared. 
\end{abstract}
\pacs{02.30.Rz 11.10.Gh 12.38.Aw 14.70.Dj \hfill  ANL-PHY-8683-TH-97,
UNITU-THEP-7/1997 }
]

A theoretical understanding of confinement of quarks and gluons into
colorless hadrons could be obtained by proving the failure of the cluster
decomposition property for color--nonsinglet gauge--covariant operators.
One long established idea in this direction is based on the occurrence of
infrared divergences to suppress the emission of colored states from
color--singlet states \cite{MP78}. Such a description of confinement in terms
of perturbation theory necessarily has to fail. 

Thus, to study the infrared behavior of QCD amplitudes non--perturbative
methods are required, and, since divergences are anticipated, a formulation in
the continuum is desirable. Both of these are provided by studies of truncated
systems of Dyson--Schwinger equations (DSEs), the equations of motion of
QCD Green's functions. Typically, for their truncation, additional sources
of information like the Slavnov--Taylor identities, entailed by gauge
invariance, are used to express vertex functions in terms of the elementary
2--point functions, {\it i.e.}, the quark, ghost and gluon propagators. Those
propagators can then be obtained as selfconsistent solutions to non--linear
integral equations representing a closed set of truncated DSEs. Some
systematic control over the truncating assumptions can be obtained
by successively including higher n--point functions in selfconsistent
calculations, and by assessing their influence on lower n--point
functions in this way. At present, even at the level of propagators no
complete solution to truncated DSEs of QCD exists. In particular, even in
absence of quarks, solutions for the gluon propagator in Landau gauge rely on
neglecting ghost contributions \cite{Man79,Atk81,Bro89,Hau96}. While this
particular problem is avoided in ghost free gauges such as the axial gauge,
in studies of the gluon DSE in this gauge \cite{BBZ81}, the
possible occurrence of an independent second term in the tensor structure of
the gluon propagator has so far been disregarded \cite{Bue95}. In fact, if
the complete tensor structure of the gluon propagator in axial gauge is taken
into account, one arrives at equations of no less
complexity than the ghost--gluon system in the Landau gauge \cite{MPpriv}.  

In addition to the prospect of some insight into con\-fine\-ment from
studying the infrared behavior of QCD Green's functions, DSEs have proved to
be a highly successful tool in developing a hadron phenomenology that
interpolates smoothly between the infrared (non--perturbative) and 
ultraviolet (perturbative) regimes \cite{Rob94}. In particular, a variety of
models for the interactions of quarks mediated by gluons exists, which are
very well suited for a dynamical description of chiral symmetry breaking from
the DSE of the quark propagator in some analogy to the gap equation in
superconductivity \cite{Miranski}. The superficial result of these studies is
that for the  quark self--energy to reflect a spontaneous breaking of chiral
symmetry  there has to be some sufficient interaction strength at low
energies. Under these circumstances, the dichotomy of the pion as a Goldstone
boson emerging from the Bethe--Salpeter equation for quark--antiquark bound
states is very well understood and explains the smallness of its mass 
as compared to all other hadrons. 

In this letter we present a simultaneous solution of a truncated set of DSEs
for the propagators of gluons and ghosts in Landau gauge. An extension to
this selfconsistent framework to include quarks dynamically is possible and
subject to further studies. The behavior of the solutions in the infrared,
implying the existence of a fixed point at a critical coupling $\alpha_c
\approx 9.5$, is obtained analytically. The gluon propagator is shown to vanish
for small spacelike momenta in the present truncation scheme. This behavior,
though in contradiction with many previous DSE studies \cite{IR-sing}, can be
partially understood from the observation that, in our present calculation, the
previously neglected ghost propagator assumes an infrared enhancement similar
to what was then obtained for the gluon.

Besides all elementary 2--point functions, {\it i.e.}, the quark, ghost and
gluon propagators, the DSE for the gluon propagator also involves the 3-- and
4--point vertex functions which obey their own DSEs. These equations involve
successively higher n--point functions. A first step towards a truncation of
the gluon equation is to neglect all terms with 4--gluon vertices. These are
the momentum independent tadpole term, an irrelevant constant which vanishes
perturbatively in Landau gauge, and explicit 2--loop contributions to the
gluon DSE. The latter are subdominant in the ultraviolet and will thus not
affect the behavior of the solutions for asymptotically high momenta. In the
infrared it has been argued that the singularity structure of the 2--loop
terms does not interfere with the one--loop terms \cite{Vac95}. Without
contributions from 4--gluon vertices (and quarks) the renormalized equation
for the inverse gluon propagator in Euclidean momentum space is given by
\cite{Nc},   
\begin{eqnarray}
  D^{-1}_{\mu\nu}(k)
    &=& Z_3 \, {D^{\hbox{\tiny tl}}}^{-1}_{\mu\nu}(k)\,
 +  g^2 N_c\, Z_1  \frac{1}{2} \int {d^4q\over (2\pi)^4}  \nonumber \\
&& \hskip -1cm 
\times \, \Gamma^{\hbox{\tiny tl}}_{\mu\rho\alpha}(k,-p,q)
  \, D_{\alpha\beta}(q) D_{\rho\sigma}(p) \, \Gamma_{\beta\sigma\nu}(-q,p,-k)
\nonumber \\
&& \hskip -1.6cm
 - \, g^2 N_c \, \widetilde Z_1 \int {d^4q\over (2\pi)^4} \; iq_\mu \,
D_G(p)\, D_G(q)\, G_\nu(q,p)
\; , \label{glDSE}
\end{eqnarray}
where $ p = k + q$, $D^{\hbox{\tiny tl}}$ and $\Gamma^{\hbox{\tiny tl}}$ are
the tree level propagator and 3--gluon vertex, $D_G$ is the ghost
propagator and $\Gamma$ and $G$ are the fully dressed 3--point
vertex functions. The equation for the ghost propagator in Landau gauge QCD,
without any truncations, is given by
\begin{eqnarray}  
D_G^{-1}(k) &=&  - \widetilde Z_3 \, k^2 \, +\, g^2 N_c\,
\widetilde Z_1  \label{ghDSE} \\
&& \hskip -1cm \times 
\int {d^4q\over (2\pi)^4} \; ik_\mu \, D_{\mu\nu}(k-q) \, G_\nu
(k,q) \, D_G(q) \; . \nonumber \end{eqnarray}
The renormalized propagators for ghosts and gluons and the renormalized
coupling are defined from the respective bare quantities by  introducing
multiplicative renormalization constants, $\widetilde Z_3 D_G := D^0_G$, $Z_3
D_{\mu\nu} :=  D^0_{\mu\nu}$ and $Z_g g := g_0$. Furthermore, $Z_1 = Z_g
Z_3^{3/2}$, $\widetilde Z_1 = Z_g Z_3^{1/2} \widetilde Z_3 $, and we use that
$\widetilde Z_1 = 1$ in Landau gauge\cite{Tay71}. The ghost and gluon
propagators are parameterized by their respective renormalization functions
$G$ and $Z$, 
\begin{equation} D_G(k) = -{G(k^2)\over k^2} ,  \;
        D_{\mu\nu}(k) = \bigg( \delta_{\mu\nu} - \frac{k_\mu
k_\nu}{k^2}\bigg) {Z(k^2)\over k^2} . 
 \label{rfD}
\end{equation}
In order to arrive at a closed set of equations for the func-
\vskip 2mm \hrule width 86mm \newpage \noindent
tions $G$ and $Z$, we use a form for the ghost--gluon vertex which is based
on a construction from its Slavnov--Taylor identity (STI) neglecting
irreducible 4--ghost correlations in agreement with the present level of
truncation \cite{ghostSTI},
\begin{equation}
 G_\mu(p,q) = \,  iq_\mu \, {G(k^2) \over G(q^2)} \, + \, ip_\mu
\, \biggl( {G(k^2)\over G(p^2)} \, - 1 \biggr) . \label{fvs} 
\end{equation}
With this result, we can construct the 3--gluon vertex according to
general procedures from previous studies \cite{3glvert},
\begin{eqnarray} 
 \Gamma_{\mu\nu\rho}(p,q,k)
       &=&  
\frac{1}{2} A_+(p^2,q^2;k^2)\,  \delta_{\mu\nu}\,  i(p-q)_\rho\,
\label{3gv} \\
&& \hskip -2cm
      + \,\frac{1}{2}  A_-(p^2,q^2;k^2)\,  \delta_{\mu\nu} \, i(p+q)_\rho
 +\, \frac{A_-(p^2,q^2;k^2)}{p^2-q^2} 
\nonumber \\
&& \hskip -2cm \times \, ( \delta_{\mu\nu} pq \, -\,  p_\nu 
q_\mu) \, i(p-q)_\rho\,  + \, \hbox{cyclic permutations} \; , \nonumber 
\end{eqnarray} 
\[   A_\pm (p^2,q^2;k^2)
   \,=\,    \frac{G(k^2) G(q^2)}{G(p^2)Z(p^2)}\, \pm
\, \frac{ G(k^2) G(p^2)}{G(q^2)Z(q^2)}  \;. \]
Some additionally possible terms, transverse with respect to all three gluon
momenta, cannot be constrained by its STI and are thus disregarded. For the
fermion vertex in QED as constructed from its Ward--Takahashi identity it is
well known that additional transverse terms, with the further constraint not 
to introduce kinematic singularities, are essential for multiplicative
renormalizability \cite{Bro91}. Based on this requirement such terms
have been obtained explicitly for quenched QED in ref.~\cite{tr3Gl}. Similar
constructions for the vertices in QCD are presently not available. However,
the full Bose (exchange) symmetry of the 3--gluon vertex alleviates this
problem since, combined with the STI, it puts much tighter constraints on
this vertex then those obtained for fermion vertices. 

Instead of a direct numerical solution of the coupled system of
integral equations resulting from the present truncation scheme we use a
one--dimensional approximation: For integration momenta $q^2 < k^2 $ we use
the angle approximation replacing $G((k-q)^2) \to G(k^2)$ and $Z((k-q)^2) \to
Z(k^2)$. Since this preserves the limit $q^2 \to 0$, it is suitable for an
analytic discussion of the solutions in the infrared. For $q^2 >
k^2$ we replace {\sl all} arguments (including the external $k^2$)
by the integration momentum $q^2$. The justification for this is the
weak logarithmic momentum dependence of $G$ and $Z$ at high momenta
\cite{AA}. The DSEs (\ref{glDSE}) and (\ref{ghDSE}) then simplify to 
\begin{eqnarray}
&& \hskip -9cm  \frac{1}{Z(k^2)}
     \,=\, Z_3 \, +\,Z_1  \frac{g^2}{16\pi^2} \Bigg\{ \int_{0}^{k^2}
\frac{dq^2}{k^2} \, \left(   \frac{7}{2}\frac{q^4}{k^4}
                     - \frac{17}{2}\frac{q^2}{k^2}
                     - \frac{9}{8} \right)  Z(q^2) G(q^2)
\,+\,  \int_{k^2}^{\Lambda_{\hbox{\tiny{UV}}}^2} 
\frac{dq^2}{q^2} \, \left( \frac{7}{8} \frac{k^2}{q^2} - 7 \right)  Z(q^2)
G(q^2) \Bigg\}   \nonumber \\
&& \hskip -7cm +\,  \frac{g^2}{16\pi^2} \Bigg\{ \int_{0}^{k^2}
\frac{dq^2}{k^2} 
\frac{3}{2} \frac{q^2}{k^2} G(k^2) G(q^2) \, - \, \frac{1}{3} G^2(k^2)
+\,\frac{1}{2} \int_{k^2}^{\Lambda_{\hbox{\tiny{UV}}}^2} \frac{dq^2}{q^2} \,
G^2(q^2) \Bigg\} \; , \label{odZDSE} \\ 
&& \hskip -8cm
 \frac{1}{G(k^2) } \,=\, \widetilde Z_3 - \frac{g^2}{16\pi^2} \, \frac{9}{4}
    \Bigg\{ \, \frac{1}{2} \, Z(k^2) G(k^2)
    \, + \,\int_{k^2}^{\Lambda_{\hbox{\tiny{UV}}}^2} \frac{dq^2}{q^2} \,
    Z(q^2) G(q^2) \Bigg\}  \; .
  \label{odGDSE}
\end{eqnarray}
\newpage
We introduced an $O(4)$--invariant momentum cutoff
$\Lambda_{\hbox{\tiny{UV}}}$ to account for logarithmic ultraviolet
divergences which are absorbed by the renormalization constants $Z_3$ and
$\widetilde Z_3$. $Z_1$ has to be ultraviolet finite \cite{us}.  
This is inconsistent with gauge invariance implying $Z_1 = Z_3/\widetilde
Z_3$. While this problem, appearing at order $g^4$ in a perturbative
expansion, is quite natural for a truncation scheme neglecting explicit
4--gluon couplings at the same order, its remedy could provide
information on purely transverse terms in the 3--gluon vertex.
For details of the renormalization and the numerical procedure see
\cite{us}. 

To deduce the infrared behavior of the propagators we make the 
Ansatz that for $x := k^2 \to 0$ the product $Z(x)G(x) \to c x^\kappa$ with
$\kappa \not= 0$ and some constant $c$. The special case $\kappa = 0$ leads
to a logarithmic singularity in eq.~(\ref{odGDSE}) for $x \to 0$ which
precludes the possibility of a selfconsistent solution. In order to obtain a
positive definite function $G(x)$ for positive $x$ from an equally positive
$Z(x)$, as $x\to 0$, we obtain the further restriction $ 0 <  \kappa <
2 \,$. Eq.~(\ref{odGDSE}) then yields,  
\begin{eqnarray}  
  G(x) &\to &  \left( g^2\gamma_0^G \left(\frac{1}{\kappa} - \frac{1}{2}
\right) \right)^{-1}  c^{-1} x^{-\kappa} \quad \Rightarrow   \label{loirG} \\
Z(x) & \to &  \left( g^2\gamma_0^G \left(\frac{1}{\kappa} - \frac{1}{2}
\right) \right) \,  c^{2} x^{2\kappa}   \; ,\label{loirZ1} \end{eqnarray}
where $\gamma_0^G = 9/(64\pi^2) $ is the leading perturbative
coefficient of the anomalous dimension of the ghost field. Using
(\ref{loirG}) and (\ref{loirZ1}) in eq. (\ref{odZDSE}), we find that the
3--gluon loop contributes terms $\sim  x^\kappa $ to the gluon equation for $x
\to 0 $ while the dominant (infrared singular) contribution $\sim  x^{-
2\kappa} $ arises from the ghost--loop, {\it i.e.},  
\[
Z(x) \to  g^2\gamma_0^G \, \frac{9}{4} \left(\frac{1}{\kappa} -
\frac{1}{2}   \right)^2 \! \left( \frac{3}{2}\, \frac{1}{2-\kappa} -
\frac{1}{3} + \frac{1}{4\kappa} \right)^{-1}\!\! c^2 x^{2\kappa} . 
\]
Comparing this to (\ref{loirZ1}) we obtain a quadratic equation with a unique
solution $\kappa = (61-\sqrt{1897})/19 \simeq 0.92$ for the
exponent $\kappa < 2 \,$. The leading behavior of the gluon and ghost
renormalization functions 
is entirely due to ghost contributions. The details of the approximations to
the 3--gluon loop have no influence on these considerations. In particular,
additional transverse terms of the 3--gluon vertex, free of kinematical
singularities, will yield contributions that are even further suppressed in
the infrared. 
Compared to the Mandelstam approximation, in which the  3--gluon loop alone
determines the infrared behavior of the gluon propagator and the running
coupling in Landau gauge \cite{Man79,Atk81,Bro89,Hau96}, this shows the
importance of ghosts. The result presented here implies an infrared stable
fixed point in the non--perturbative running coupling of our subtraction
scheme, defined by 
\begin{equation}
  \alpha_S(s) = \frac{g^2}{4\pi} Z(s) G^2(s)
    \to \frac{16\pi}{9} \left(\frac{1}{\kappa} - \frac{1}{2}\right)^{-1} \!\!
    \approx 9.5 \; , 
\end{equation} 
for $s\to 0$. This is qualitatively different from the infrared singular 
coupling of the Mandelstam approximation \cite{Hau96}.

The momentum scale in our calculations is fixed from the phenomenological
value $\alpha_S (M_Z) = 0.118$ at the mass of the $Z$--boson \cite{PDG96}.
The ratio of the $Z$-- to the $\tau$--mass, $M_Z/M_\tau \simeq 51.5$, then
yields $\alpha_S(M_\tau) = 0.38$ which is in encouraging agreement with
the experimental value.

\begin{figure}
  \centering{\
        \epsfig{file=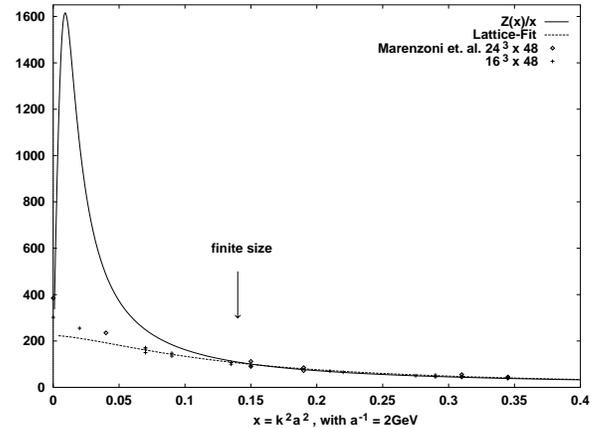,height=0.65\linewidth,width=0.94\linewidth}}
\vskip .5cm
\caption{The numerical result for the gluon propagator from Dyson--Schwinger
         equations (solid line) compared to lattice data from fig.~3 in
         \protect\cite{Mar95}.}
  \label{Zlat}
\end{figure}

It is interesting to compare our solutions to recent lattice results
using implementations of the Landau gauge condition
\cite{Ber94,Mar95,Sum96}. In figure \ref{Zlat} we compare our solution for
the gluon propagator to 
data from ref.~\cite{Mar95}. We normalized the gluon propagator according to
$Z(x = 1) \approx 11.3$ to account for the units used in ref.~\cite{Mar95}
(with $x = k^2 a^2 $ in units of the inverse lattice spacing). According to
the authors of \cite{Mar95}, the arrow indicates a bound below which finite
size effects become considerable. 

\begin{figure}
  \centering{\
        \epsfig{file=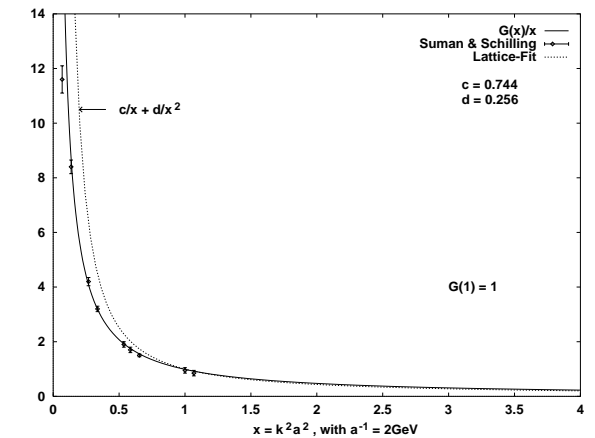,height=0.65\linewidth,width=0.94\linewidth}}
\vskip 5mm
\caption{The numerical result for the ghost propagator from Dyson--Schwinger
         equations (solid line) compared to data from fig.~1 in
         \protect\cite{Sum96} for the $24^4$ lattice up to $x \approx 1$,
         and a fit as obtained in ref.\ \protect\cite{Sum96} for $x \ge 2$.} 
  \label{Glat}
\end{figure}

In figure \ref{Glat} we compared our infrared enhanced ghost propagator to
the results of ref. \cite{Sum96}. It is quite amazing to observe that
our solution fits the lattice data at low momenta significantly
better than the fit to an infrared singular form $D_G(k^2) = c/k^2 + d/k^4$
given in \cite{Sum96}. We therefore conclude that present lattice
calculations confirm the existence of an infrared enhanced ghost propagator
of the form $D_G \sim 1/(k^2)^{1+\kappa}$ with $0 < \kappa < 1$. This is an
interesting result for yet another reason: In ref.~\cite{Sum96} the 
Landau gauge condition was supplemented by an algorithm to select 
gauge field configurations from the fundamental modular region which is
to eliminate systematic errors that might occur due to the presence of 
Gribov copies. Thus, our results suggest that the existence of such
copies of gauge configurations might have little effect on the solutions to
Landau gauge DSEs \cite{Zwa94}. 

The Euclidean gluon correlation function presented here can be shown to
violate reflection positivity \cite{us}, which is a necessary and sufficient
condition for the existence of a Lehmann representation \cite{Fritz}. We
interpret this as representing confined gluons. In order to understand how
these correlations can give rise to confinement of quarks, it will be necessary
to include the quark propagator. The size of the coupling at the fixed point,
$\alpha_c \approx 9.5$, is however, a good indication that dynamical chiral
symmetry breaking will be generated in the quark DSE. 

In summary, we presented a solution to a truncated set of coupled
Dyson--Schwinger equations for gluons and ghosts in Landau gauge. The
infrared behavior of this solution, obtained analytically, represents a
strongly infrared enhanced ghost propagator and an infrared vanishing gluon
propagator. 
Our results, in particular for the ghost propagator, compare favorably with
recent lattice calculations \cite{Mar95,Sum96}. Since the lattice
implementations of the Landau gauge are such that configurations are
restricted to the fundamental modular region, this might indicate that Gribov
copies have little influence on solutions to the DSEs in Landau gauge. The
absence of a Lehmann representation for the gluon propagator can be
interpreted as signal for confined gluons. The existence of an infrared fixed
point is in qualitative disagreement with previous studies of the gluon DSE
neglecting ghost contributions in Landau gauge
\cite{Man79,Atk81,Bro89,Hau96}. This shows that ghosts are important, in
particular, at low energy scales relevant to hadronic observables. 

We thank F.~Coester, F.~Lenz, M.~R.~Pennington and H.~Reinhardt for helpful
discussions. This work was supported by DFG under contract Al 279/3-1, by
the Graduiertenkolleg T\"ubingen  and the US-DOE, Nuclear Physics Division,
contract \#\ W-31-109-ENG-38.

\begin{thebibliography}{99}
\bibitem{MP78}  W.~Marciano and H.~Pagels, Phys.~Rep.\ {\bf 36}, 137 (1978).
\bibitem{Man79} S.~Mandelstam, Phys.~Rev.~D {\bf 20}, 3223 (1979).
\bibitem{Atk81} D.~Atkinson et al.,
                J.~Math.~Phys.~{\bf 22}, 2704 (1981);
                D.~Atkinson, P.~W.~Johnson and K.~Stam,
                {\it ibid.} {\bf 23}, 1917 (1982).
\bibitem{Bro89} N.~Brown and M.~R.~Pennington,
                Phys.~Rev.~D {\bf 39}, 2723 (1989).
\bibitem{Hau96} A.~Hauck, L.~v.~Smekal and R.~Alkofer, {\it e-print},
                hep-ph/9604430
\bibitem{BBZ81} M.~Baker, J.~S.~Ball and F.~Zachariasen, Nucl.~Phys.\
                {\bf B186}, 531/560 (1981); W.~J.~Schoenmaker,
                {\it ibid.} {\bf B194}, 535 (1982);  J.~R.~Cudell and
                D.~A.~Ross, {\it ibid.} {\bf B358}, 247 (1991).
\bibitem{Bue95} K.~B\"uttner and M.~R.~Pennington, Phys.~Rev.~D {\bf 52},
                5220 (1995).
\bibitem{MPpriv} R.~Alkofer, M.~R.~Pennington, L.~v.~Smekal and P.~Watson,
                 work in progress.
\bibitem{Rob94} See, {\it e.g.}, C.~D.~Roberts and A.~G.~Williams, Prog.\
                Part.\ Nucl.~Phys.\ {\bf 33}, 477 (1994), and references
                therein.
\bibitem{Miranski}
                See also ``Dynamical Symmetry Breaking in Quantum Field
                Theories'', V.~A.~Miranski, World Scientific, 1993.
\bibitem{IR-sing} An infrared enhanced gluon propagator was found in Landau
                  gauge in Mandelstam approximation
                  \protect\cite{Man79,Atk81,Bro89,Hau96} as well as in some
                  studies of its simplified axial gauge DSE
                  \protect\cite{BBZ81,Bue95}.
\bibitem{Vac95} L.~G.~Vachnadze, N.~A.~Kiknadze and A.~A.~Khelashvili,
                Theor.~Math.~Phys.~{\bf 102}, 34 (1995).
\bibitem{Nc}    We use positive definite metric, $g_{\mu\nu} =
                \delta_{\mu\nu}$. Color indices are suppressed and the number
                of colors is fixed, $N_c = 3$.  
\bibitem{Tay71} J.~C.~Taylor, Nucl.~Phys.~{\bf B33}, 436 (1971).
\bibitem{ghostSTI} In \protect\cite{us} we derive a Slavnov--Taylor identity
                   for the ghost--gluon vertex from the usual BRS invariance.
                   This together with the symmetry of the ghost--gluon vertex
                   fully determines its form at the present level of
                   truncation. There are no undetermined transverse terms
                   in this case. 
\bibitem{3glvert} The ghost--gluon vertex (\protect\ref{fvs}) is consistent
                  with a ghost--gluon scattering kernel of tree--level
                  structure in the STI of the 3--gluon vertex, which can then
                  be constructed from procedures developed and used
                  previously (U. Bar--Gadda, Nucl. Phys. {\bf B163}, 312 (1980;
                  S.~K.~Kim and M.~Baker, {\it ibid.} {\bf B164}, 152 (1980);
                  J.~S.~Ball and {T.-W.~Chiu}, Phys.~Rev.~D {\bf 22}, 2550
                  (1980)). 
\bibitem{Bro91} N. Brown and N. Dorey, Mod. Phys. Lett. {\bf A6},
                317 (1991).
\bibitem{tr3Gl} D. C. Curtis and M. R. Pennington, Phys. Rev. D {\bf 42}, 4165
                (1990).
\bibitem{AA}    A similar assumption underlies the Mandelstam approximation.
                For $q^2 > k^2$ it can be further justified from a detailed
                study of the solutions in the ultraviolet \protect\cite{us}.
                It will nevertheless be important to assess the sensitivity of
                the results to the modified angle approximation in future. 
\bibitem{us}    L.~v.~Smekal, A.~Hauck and R.~Alkofer, {\it e--print},
                hep--ph/9707327.
\bibitem{PDG96} Particle Data Group, Phys.~Rev.~D {\bf 54}, 77 (1996).       
\bibitem{Ber94} C.~Bernard, C.~Parrinello and A.~Soni, Phys.~Rev.~D {\bf 49},
                1585 (1994); D.~S.~Henty et al., {\it ibid.} {\bf 54} (1996),
                6923.  
\bibitem{Mar95} P.~Marenzoni, G.~Martinelli, and N.~Stella, Nucl.~Phys.\ 
                {\bf B455}, 339 (1995).
\bibitem{Sum96} H.~Suman and K.~Schilling, Phys.~Lett.~{\bf B373}, 314 (1996).
\bibitem{Zwa94} This is supported by the qualitative similarity of our
                solutions to the infrared behavior obtained from 
                studies of the influence of a complete gauge fixing by 
                D.~Zwanziger, Nucl.~Phys. {\bf B378}, 525 (1992); {\it ibid.}
                {\bf B412}, 657 (1994).
\bibitem{Fritz} We are indebted to F.~Coester for pointing this out. 
\end{thebibliography}
\end{document}